\documentstyle[12pt]{article}
\begin{document}
\begin{center}
{\Large \bf
The extra high energy cosmic rays spectrum
in view of the decay of proton into Planck neutrinos
}
\bigskip

{\large D.L.~Khokhlov}
\smallskip

{\it Sumy State University, R.-Korsakov St. 2\\
Sumy 244007 Ukraine\\
e-mail: khokhlov@cafe.sumy.ua}
\end{center}

\begin{abstract}
It is assumed that proton decays into Planck neutrinos.
The energy of the Planck neutrino is equal to the Planck mass
in the preferred rest frame.
In the frame moving relative to the preferred rest frame
the energy of the Planck neutrino reduces by
the corresponding Lorentz factor.
The lifetime of proton depends on the kinetic energy
of proton relative to the preferred rest frame.
The time required for proton travel from the source to the
earth defines the limiting energy.
Protons with the energies more than the limiting energy decay and
do not give contribution in the EHECRs spectrum.
It is shown that EHECRs with the energies
$E>3 \times 10^{18}\ {\rm eV}$ can be identified with
the Planck neutrinos.
\end{abstract}

The energy spectrum of extra high energy cosmic rays
(EHECRs) above $10^{10}\ {\rm eV}$ can be
divided into three regions: two "knees" and one "ankle"~\cite{Yo}.
The first "knee" appears around $3\times 10^{15}\ {\rm eV}$
where the spectral power law index changes from $-2.7$ to $-3.0$.
The second "knee" is somewhere between $10^{17}\ {\rm eV}$ and
$10^{18}\ {\rm eV}$ where the spectral slope steepens from
$-3.0$ to around $-3.3$. The "ankle" is seen in the region of
$3 \times 10^{18}\ {\rm eV}$ above which the spectral slope
flattens out to about $-2.7$.

It was proposed~\cite{Kh3} that, at the Planck scale
$m_{Pl}=1.2 \times 10^{19}\ {\rm GeV}$, the decay of electron
into neutrino (Planck neutrino) occurs
\begin{equation}
e\rightarrow \nu_{Pl}.
\label{eq:enu}
\end{equation}
Within the framework of electrodynamics,
hadrons can be described by the structure of 5 electrons~\cite{Kh2}.
The structure of proton is given by
\begin{equation}
p\equiv e^+e^-e^+e^-e^+.
\label{eq:pro}
\end{equation}
From this, at the Planck scale, proton decays into 5 Planck neutrinos.
The lifetime of proton relative to the decay
into Planck neutrinos is given by
\begin{equation}
t_p=t_{Pl}\left(\frac{m_{Pl}}{2m_p}\right)^5
\label{eq:tp1}
\end{equation}
where the factor 2 takes into account the transition from
the massive particle to the massless one.

The energy of the Planck neutrino depends on the reference frame.
Let, in the preferred rest frame,
the energy of the Planck neutrino be equal to the Planck mass
\begin{equation}
E_{\nu}=m_{Pl}.
\label{eq:Enu}
\end{equation}
In the frame moving relative to the preferred rest frame
with the Lorentz factor
\begin{equation}
\gamma=\left(1-\frac{v^2}{c^2}\right)^{1/2}
\label{eq:gam}
\end{equation}
the energy of the Planck neutrino reduces by
the Lorentz factor
\begin{equation}
E_{\nu}'=\gamma m_{Pl}.
\label{eq:Enu1}
\end{equation}
From this the lifetime of proton in
the frame moving relative to the preferred rest frame
is given by
\begin{equation}
t_p=t_{Pl}\left(\frac{\gamma m_{Pl}}{2m_p}\right)^5.
\label{eq:tp2}
\end{equation}

In view of eq.~(\ref{eq:tp2}), the lifetime of proton decreases
with the increase of the kinetic energy of proton
relative to the preferred rest frame given by
\begin{equation}
E_{p}=\frac{m_{p}}{\gamma}.
\label{eq:Ep}
\end{equation}
Let the earth possess unity Lorentz factor
relative to the preferred rest frame.
For protons arrived at the earth,
the travel time meets the condition
\begin{equation}
t\leq t_p.
\label{eq:t}
\end{equation}
From this the time required for proton travel from the source to the
earth defines the limiting energy of proton
\begin{equation}
E_{lim}=\frac{m_{Pl}}{2}\left(\frac{t_{Pl}}{t}\right)^{1/5}.
\label{eq:E}
\end{equation}
Within the time $t$, protons
with the energies $E>E_{lim}$ decay and
do not give contribution in the EHECRs spectrum.

Determine
the range of the limiting energies of proton
depending on the range of distances to the EHECRs sources.
Take the maximum and minimum distances to the source as
the size of the universe and the thickness
of our galactic disc respectively.
For the lifetime of the universe
$t_U=1.06 \times 10^{18}\ {\rm s}$~\cite{Kh1},
the limiting energy is equal to $E_U=3.3 \times 10^{15}\ {\rm eV}$.
This corresponds to the first "knee" in the EHECRs spectrum.
For the thickness of our galactic disc $\simeq 300\ {\rm pc}$,
the limiting energy is equal to $E_G=5.5 \times 10^{17}\ {\rm eV}$.
This corresponds to the second "knee" in the EHECRs spectrum.
Thus
the range of the limiting energies of proton
due to the decay of proton into Planck neutrinos
lies between
the first "knee" $E\sim 3\times 10^{15}\ {\rm eV}$ and
the second "knee" $E\sim 10^{17}-10^{18}\ {\rm eV}$.

From the above consideration it follows that
the decrease of the spectral power law index from $-2.7$ to $-3.0$
at the first "knee" $E\sim 3\times 10^{15}\ {\rm eV}$ and
from $-3.0$ to around $-3.3$
at the second "knee" $E\sim 10^{17}-10^{18}\ {\rm eV}$
can be explained as a result of
the decay of proton into Planck neutrinos.
From this it seems natural that, below
the "ankle" $E<3 \times 10^{18}\ {\rm eV}$,
the EHECRs events are mainly caused by the protons.
Above the "ankle" $E>3 \times 10^{18}\ {\rm eV}$,
the EHECRs events are caused by the particles other than protons.

If Planck neutrinos take part in the strong interactions,
they must give contribution in the EHECRs events.
Since proton decays into 5 Planck neutrinos,
the energy of the Planck neutrino is $1/5$ of the energy of
the decayed proton.
For the spectral power law index equal to $-2.7$,
the ratio of the proton flux to the Planck neutrino flux
is given by
$J_p/J_{\nu}=5^{1.7}=15$.

From the above consideration it is natural to identify EHE
particles with the energies $E>3 \times 10^{18}\ {\rm eV}$
with the Planck neutrinos.
Continue the curve with the spectral power law index $-2.7$
from the "ankle" $E\sim 3 \times 10^{18}\ {\rm eV}$ to
the first "knee" $E\sim 3\times 10^{15}\ {\rm eV}$ and
compare the continued curve with the observational curve.
Comparison gives
the ratio of the proton flux to the Planck neutrino flux
$J_p/J_{\nu}=15$.

\end{document}